\documentclass[notitlepage,nofootinbib, superscriptaddress, 12pt,
  a4paper]{revtex4-1}

\usepackage{graphicx}
\usepackage{dcolumn}%
\usepackage{bm}
\usepackage{xspace}
\usepackage{amsmath,amsfonts,amssymb} 
\usepackage[colorlinks]{hyperref}
\usepackage{color} 
\usepackage{hyperref} 
\usepackage[normalem]{ulem}

\newcommand{\mev}[1]{\ensuremath{#1 \text{\,MeV}}\xspace}
\newcommand{\gevc}[1]{\ensuremath{#1 \text{\,GeV/$c$}}\xspace}

\newcommand{\pp}{p-p\xspace} 
\newcommand{\ppb}{p-Pb\xspace}

\newcommand{\ee}{\ensuremath{e^+e^-}\xspace}

\newcommand{\Tc}{\ensuremath{T_{\text{c}}}\xspace}

\newcommand{\qs}{\ensuremath{Q_{\text{s}}}\xspace}
\newcommand{\qd}{\ensuremath{Q_{\text{d}}}\xspace}
\newcommand{\qh}{\ensuremath{Q_{\text{h}}}\xspace}
\newcommand{\dndeta}{\ensuremath{dN/d\eta}\xspace}

\newcommand{\as}{\ensuremath{\alpha_{\text{s}}}\xspace}
\newcommand{\lqcd}{\ensuremath{\Lambda_{\text{QCD}}}\xspace}

\newcommand{\snnt}[1]{\ensuremath{\sqrt{s_{\text{NN}}} = #1\,\text{TeV}}\xspace}
\newcommand{\et}{\ensuremath{E_{\rm T}}\xspace}

\begin{document}

\title{What Quark-Gluon Plasma in small systems\\
might tell us about nucleons}

\author{P. Christiansen} 
\affiliation{Division of Particle Physics, Lund University, Sweden}
\date{\today}

\begin{abstract}
The origin of flow-like effects in small systems, such as those produced in
ultra-relativistic proton-proton and proton-lead collisions, is still widely
debated. In this paper the goal is to look at possible consequences if indeed
a mini-Quark-Gluon Plasma is formed in these collisions. It is argued that
this could indicate a duality between the QGP phase and the color fields in
hadrons. A qualitative dense field picture is presented for this duality and
discussed.
\end{abstract}

\maketitle

\section{Introduction}
\label{sec:intro}

There are indications that a Quark-Gluon Plasma (QGP) is produced even in \pp
and \ppb
collisions~\cite{Khachatryan:2010gv,Abelev:2012ola,ABELEV:2013wsa,Aaboud:2016yar}. There
is a lot of theoretical and experimental work still to be done, but the goal
of this paper is to start speculating on possible implications beyond the
collisional regime of mini QGPs. In this paper, it is therefore assumed that a
mini QGP is created in each \pp collision, meaning that there is a soft
underlying part of the event that forms a QGP (hard jets are not QGP like, but
will behave more or less like standard jets in \ee collisions).

The paper is centered around 2 key ideas. In the first part,
Sec.~\ref{sec:qgp_large} and \ref{sec:small}, it is argued based on QGP
properties in large systems that time reversal could be a good symmetry
between the QGP and hadronic state. In small systems, such a symmetry would
suggest a duality between QGP and hadrons. In the second part,
Sec.~\ref{sec:duality}, a qualitative model of such a duality is
explored. Finally, in Sec.~\ref{sec:discussion} the ideas of the paper is
discussed in a broader context.

\section{The QGP properties in large systems}
\label{sec:qgp_large}

Traditionally, small systems have been thought of as the baseline for
understanding large systems, e.g., jet quenching and quarkonium
melting. However, for bulk effects in small systems it appears that large
systems is the baseline because the unexpected effects are easier to identify
and isolate there. Here, we shall take the idea a step further and assume that
the QGP ``standard'' model developed for large systems is a baseline for the
QGP formation also in small systems\footnote{These ideas are all fairly
  standard, see e.g., ``NuPECC Long Range Plan 2016/2017'', therefore few
  references are given.}

A heavy-ion collision proceeds through the following stages:
\begin{enumerate}
\item Initial scatterings and QGP formation
\item QGP expansion
\item Hadronization
\item Chemical freeze out
\item Kinetic freeze out
\end{enumerate}

The initial scatterings and QGP formation are the least understood and as this
will not play a role for the discussion here, this step is skipped.

The QGP phase behaves like a nearly perfect liquid. The expansion of a perfect
liquid is reversible, meaning that the expansion, as the QGP cools, generates
essentially no entropy (the relative entropy increase is as small as it can
be). This behavior is expected to be relatively constant in a reasonable large
temperature range above \Tc, meaning that the QGP behaves as a nearly perfect
liquid in the whole phase transition region.

The transition from QGP to hadrons is a crossover transition so it is not a
real phase transition. There is no sharp separation between the two phases,
instead they coexist in some temperature range around the pseudo-critical
temperature, \Tc. In such a crossover transition there is no entropy
generation and so the change between hadronic and partonic degrees
of freedom is fast.

On the hadronic side, statistical thermal models have been very successful at
describing the particle composition with temperatures similar to \Tc
(e.g.,~\cite{Stachel:2013zma}), and there are so far no indication that the
hadronic states at \Tc are different from those at lower temperatures. A
significant amount of hadrons produced are resonances that decay. These decays
are not reversible.

Finally, there have been some model results that suggests that both the
chemical and kinetic freeze out occurs essentially at \Tc,
e.g.~\cite{Broniowski:2001uk}. At the LHC, resonance results indicates effects
of hadronic rescattering between the chemical and thermal freeze
out~\cite{Abelev:2014uua}. The effect must be less for small systems and is
not critical for the ideas here, which focus on the hadronization, so we will
not discuss this in any more detail here.

It might seem counterintuitive that large features of these collisions are
reversible, but one should recall that this is exactly what allows us to probe
the initial state geometry using the final state flow.

\section{What do we learn about QCD if the QGP is formed in small system}
\label{sec:small}

Now let us try to consider dilute \pp collisions. The immediate problem is
that we do not understand the initial scatterings and QGP formation process
and so it can seem hard to pin down exactly what we learn from observing the
QGP in these systems. However, as we have seen in the previous section a lot
of features in large systems are reversible, which makes it interesting to
consider time reversal. If time reversal is applicable, we can get from the
final state to the QGP phase through stages that are reasonable well
understood as described in the previous section. In this section it will be
argued that, if our assumption of large-system-like QGP formation in small
systems are true, then time reversal is a good symmetry in small systems. This
feature will be used to highlight that QGP formation in small systems in fact
suggests that this is not a high energy feature but a general feature of
essentially all hadronic collisions.

Since the assumption is that QGP is produced in the underlying event of all \pp
inelastic collisions then we can select the simplest to analyze. This means we
can suppress the influence of jets and rare events where high mass
resonances are produced. We can also focus on dilute \pp collisions where few
(tens of) particles are produced in a wide rapidity region around midrapidity. In these dilute collisions we expect final state hadronic
rescattering to play an even smaller role than in the previous section and so
we can ignore the kinetic freeze out phase. Furthermore, in these dilute
collisions the bulk of the initial kinetic energy of the colliding protons
remains at very forward rapidities as fragments that are decoupled from the
QGP evolution and so we can neglect them as we just want to go back into the
QGP phase. So it is now just the few low-energy particles at midrapidity that
we want to use time reversal on to point out that they would reform a QGP if
they were recollided by inverting their momentum vectors.

In the time reversal argument we, following the arguments in the previous
paragraph, start at chemical freeze out. Resonance decays are non-reversible
but as the freeze-out temperature is significantly smaller than the Hagedorn
temperature there is no indication that hadron production is dominated by
decays of very heavy resonances. Instead it seems likely that we can select
\pp collisions where this is not the case. For these collisions, time reversal
takes us back through the phase transition and well into the QGP phase
itself. What this reversibility exactly means at the Quantum Mechanical level
is a bit unclear to the author. Here, the point is not if it evolves back to
the exact same state but that it evolves back to a state with similar
properties (a QGP).

So, if we do not observe an experimental size threshold for QGP formation then
we can use time reversal arguments to argue that there is not really an energy
threshold for when the QGP is created - the QGP formation and inelastic
thresholds are the same! However, this does not mean that the effects will be
easy to observe. In low energy collisions, energy, momentum, and
quantum-number conservation could easily hide most of the effects. Another
important thing to point out is that even if we at LHC could observe
ridge-like effects down to a few particles, as, e.g., ATLAS is
pursuing~\cite{Aaboud:2016yar}, then the origin of the ridge in the perfect
liquid picture is the initial state geometry, which spans several units of
rapidity. In low-energy collision there is therefore no reason to expect a
ridge to appear even the dynamics of the bulk matter is the same because the
initial state preparation is very different (and might not even span several
units of rapidity). \\ The lack of an energy threshold also suggests that
there will be QGP effects in \ee collisions (after the jets have fragmented)
but that these effects are likely hidden by the initial configuration of the
system, which is jet driven. To observe these effects one would probably have
to look at midrapidity in the thrust system and suppress events with hard
radiation.

Let us try to summarize the logic of this section as this is a central idea of
the paper. Assuming that a QGP is formed in all \pp collisions we can select
particular simple collisions where non-reversible physics such as resonance
decays and elastic scatterings after hadronization is negligible. We can then
go reversibly back from the final state at freeze-out and into the QGP
phase. The energetic forward going particles are only used for creating the
QGP so for dilute events we conclude that just recolliding the few low energy
hadrons produced at midrapidity would create a QGP. But if there is no size
or energy threshold for QGP formation then it suggests that the QGP is in some
sense (to be explored in the next section) present in the hadrons themselves
(as opposed to being created and only exist for a short time in high-energy
hadronic collisions).

\section{Duality between QGP and hadrons}
\label{sec:duality}

A fundamental problem in the study of the strong interaction is that QCD
describes quarks and gluons while we observe hadrons. If the origin of
QGP-like effects in small and large systems is the same, then it would mean
that heavy-ion collisions allow the direct study of fundamental QCD dynamics
that is difficult to isolate in small systems. This picture is completely
opposite to the old idea of parton-hadron duality because it is a lump of
dense partonic medium that is now dual to a (dense) hadron. It therefore seems
motivated to propose a QGP-hadron duality. The goal in the rest of the section
is to explore a concrete idea for the proposed duality.\\

Let us first here discuss what such a duality must contain. Since we know that
the QGP is strongly interacting the duality must mean that the quarks and
gluons inside hadrons are strongly interacting. That is different from the
idea of, e.g., the Bag model where the quarks and gluons are a weakly coupled
system that is confined by the negative vacuum pressure. As the system is
strongly interacting it must be dense in terms of gluons because the strong
force needs antiscreening to be strong. It is known that the low $x$ region of
the hadronic structure is dense, which have given rise to the so called Color
Glass Condensate (CGC)~\cite{Gelis:2010nm}. The CGC gives a universal
description of nucleon and nuclei wave functions for $x \leq 0.01$ and $Q >
\gevc{1}$. If the proton is dense in general then we propose that the CGC is
one limit of a more general dense field description. \\ The task at hand is
therefore to come up with a general model for \emph{strongly interacting}
degrees of freedoms that can both describe the QGP and hadrons and can be
perceived as a generalization of the CGC degrees of freedom. We will therefore
propose that these degrees of freedoms are dense fields or dense
domains. Dense fields mean here that the relevant degrees of freedom are
coherent ``high occupancy'' color fields of a certain size, $1/\qd$ (d for
domain), as opposed to point-like quarks and gluons. For example, it will be
argued that the proton at rest is made up of 3 such dense fields.

\subsection{The origin of the dense fields}

The origin of the dense fields is assumed to be due to the antiscreening in
QCD: when a bare color charge is put in the vacuum, the color field will
physically grow into the dense field because of the antiscreening. This
growth gives rise to the dense fields in the hadrons and in the QGP,
and it is the push to extend the fields that leads to the
hydrodynamic expansion.

To understand when this growth stop (the hadron size), the energy in one of
these dense fields has to be estimated. We expect that the kinetic energy will
grow as \qd, i.e., the smaller the field the more kinetic energy it will
contain. Antiscreening in this picture reflects the density of color charge
in the dense field and so we assume that the energy stored in the field will
therefore also grow as $\as(\qd)$, so that the total energy of the dense field
is:
\begin{eqnarray}
\label{eq:energy}
E & \propto & \as(\qd) \qd \\
  & \propto & \frac{\qd}{\log{\left(\frac{\qd}{\lqcd}\right)}}, 
\end{eqnarray}
where $\lqcd \approx \mev{200}$.\\
This expression has a minimum, $\qh$, when.
\begin{equation} 
\qh = e \lqcd \approx \mev{544},
\end{equation} 
where \qh denotes the characteristic size of the dense fields in hadrons.

\begin{figure}[htbp]
  \begin{center}
   \includegraphics[width=0.7\columnwidth]{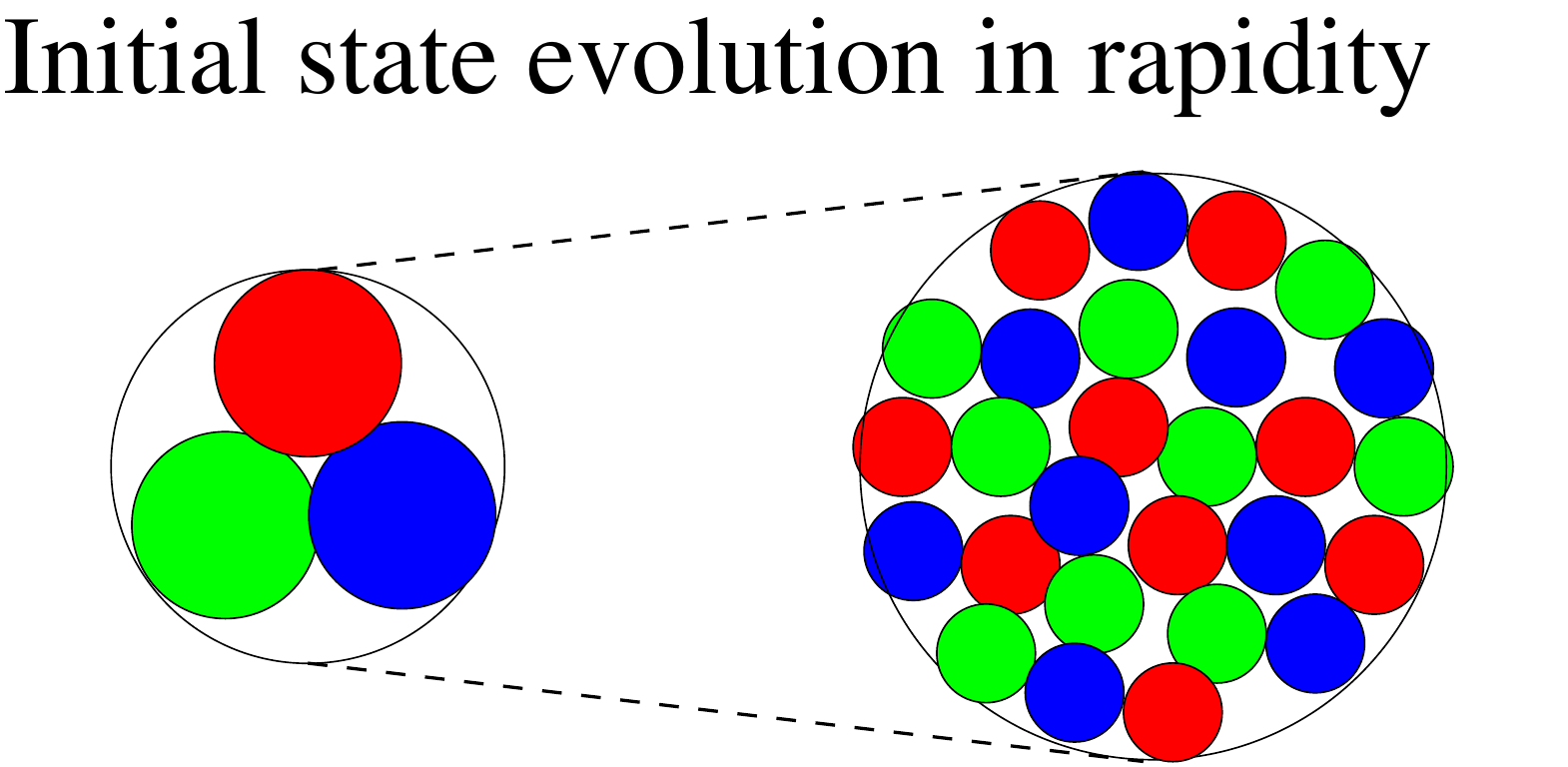}
    \caption{Initial state evolution of the dense fields in rapidity --
      similarly to the CGC picture. The figure is meant as a schematic
      illustration. Importantly, the fields in hadrons are assumed to be dense
      (no empty regions) and can have varying density (domain size) and shape.}
  \label{fig:initial}
\end{center}
\end{figure}

This means that hadrons will be made up of valence fields with the
characteristic size $1/\qh$ (up to several small factors). In particular the
proton will be made up of 3 such fields. Now we want to motivate that this
dense field description for a proton at rest can be evolved into the CGC. The
word motivate is used because there are several issues, e.g., the proton at
rest is a 3D non-perturbative object while the CGC is a 2D perturbative
object.

The basic HERA inspired CGC saturation scale is
\begin{equation}
  \label{eq:cgc}
  \qs^2(x) = 1\,\text{GeV}^2 \left( \frac{3 \cdot 10^{-4}}{x} \right)^{0.29}.
\end{equation}
The goal is now to show that, if the relative
evolution with $x$,
\begin{equation}
  \label{eq:evolution}
  \qs^2(x) \propto \left( \frac{3 \cdot 10^{-4}}{x} \right)^{0.29},
\end{equation}
 is assumed, one obtains a similar value of the absolute scale in the dense
 field picture. For the proton at rest there are 3 fields, which shares the
 mass so that $x=1/3$ seem reasonable and therefore $\qd^2(1/3) =
 \qh^2$. With this constraint and Eq.~\ref{eq:evolution} one obtains
\begin{equation}
  \qd^2(x) \approx 2.26\,\text{GeV}^2 \left( \frac{3 \cdot 10^{-4}}{x} \right)^{0.29},
\end{equation}
which is similar to Eq.~\ref{eq:cgc}. Both $\qd^2(x)$ and \qh seem too large,
which suggest that an additional common small factor is missing. The dilute
image of the proton, which we know from hard scatterings, $Q \gg \qd$, should
then be recovered by evolution in $Q$.\\

Let me shortly here discuss how this model compares to the Bag model and the
CGC. The arguments leading to Eq.~\ref{eq:energy} are reminiscent of the
arguments in Bag models with the Bag pressure replaced by the field strength
\as, but as stressed in the introduction to this section, the Bag model is a
weakly coupled picture while the dense field picture is a strongly interacting
model. We finally note that while the vacuum Bag pressure makes sense from
Quantum Mechanical arguments there exists no quantitative understanding of the
vacuum pressure of the Universe and therefore this pressure might not be
confining the hadrons.

The CGC is a model for the initial state of hadrons in high-energy hadronic
collisions~\cite{Gelis:2010nm}. It is in principle derived from first
principles, but in reality several approximations have to be done to obtain
results. The main idea is that bulk production is driven by semi-hard
interactions of low $x$ gluon fields. The gluon densities at low $x$ can
become very large for momentum scales smaller than a saturation scale $Q_s$,
so that they can be treated as classical fields. A semi-hard approach can be
taken when $Q_s$ is so large that $\alpha_s(Q_s)$ is small, which one expects
at LHC where $Q_s$ is estimated to be of order a few GeV. The CGC when
combined with PYTHIA fragmentation can reproduce many features of small
systems~\cite{Schenke:2016lrs}. \\ It should be clear that the QGP is a real
model whereas the picture proposed here is a qualitative sketch. Still, it is
important to stress in what fundamental way this sketch differs from the CGC
model. Firstly, the CGC is is a model of the initial state and not a model of
the QGP. It is only applicable at high energies as $Q_s$ has a strong rapidity
dependence whereas the proposed initial state effect here is expected to
affect all energies. Secondly, the CGC description of the QGP-like effects in
small systems~\cite{Schenke:2016lrs} does not require a QGP, nor ideal
hydrodynamics, to describe the ridge and so there is no reversible QGP
phase. This means that the whole argument chain put forward in
Sec.~\ref{sec:small} collapses. The initial state description proposed here is
a fundamental non-perturbative picture, which, if true, would suggest that the
limits in which the CGC approximations (semi-hard and classical) is valid are
not yet dominating bulk production.\\

\subsection{The dense fields in the QGP}


In this section the goal is to explore the dense field description in the
QGP. In the QGP we assume that the dense fields eventually form and it is their
push to expand, to decrease their energy cf.\ Eq.~\ref{eq:energy}, that gives
rise to the hydrodynamic expansion.

Let us first try to understand how the domain size depends on the energy
density. While the systems formed in hadronic collisions will typically have
small momenta transversely they will have significant longitudinal
momenta. Here, it is first assumed that the system is at rest when the dense
fields are formed and then the effect of the longitudinal momentum on
observables is guesstimated. For a homogenous system the domain size, $1/\qd$,
must be the same for all domains. The density of domains, $\rho$, is then
\begin{equation}
\label{eq:density}
\rho \propto \qd^3
\end{equation}
and the internal energy (mass) of each domain is proportional to $\qd$
cf.\ Eq.~\ref{eq:energy}. The energy density, $\varepsilon$, will therefore
scale with \qd as
\begin{equation}
\label{eq:edensity}
\varepsilon \propto \qd^4.
\end{equation}
The constant of proportionality can be found for a region where $\as$ is
constant ($\qd \sim \qh$) since the same equation has to be valid for
nucleons. For the QGP one knows from Lattice QCD that $\varepsilon \propto
T^4$ so this suggests that $\qd \propto T$.

As the QGP expands and cools the domain size will increase ($\qd \rightarrow
\qh)$ and eventually
reach the hadronization size where the domains will be confined inside the
hadrons. 

\begin{figure}[htbp]
  \begin{center}
   \includegraphics[width=0.7\columnwidth]{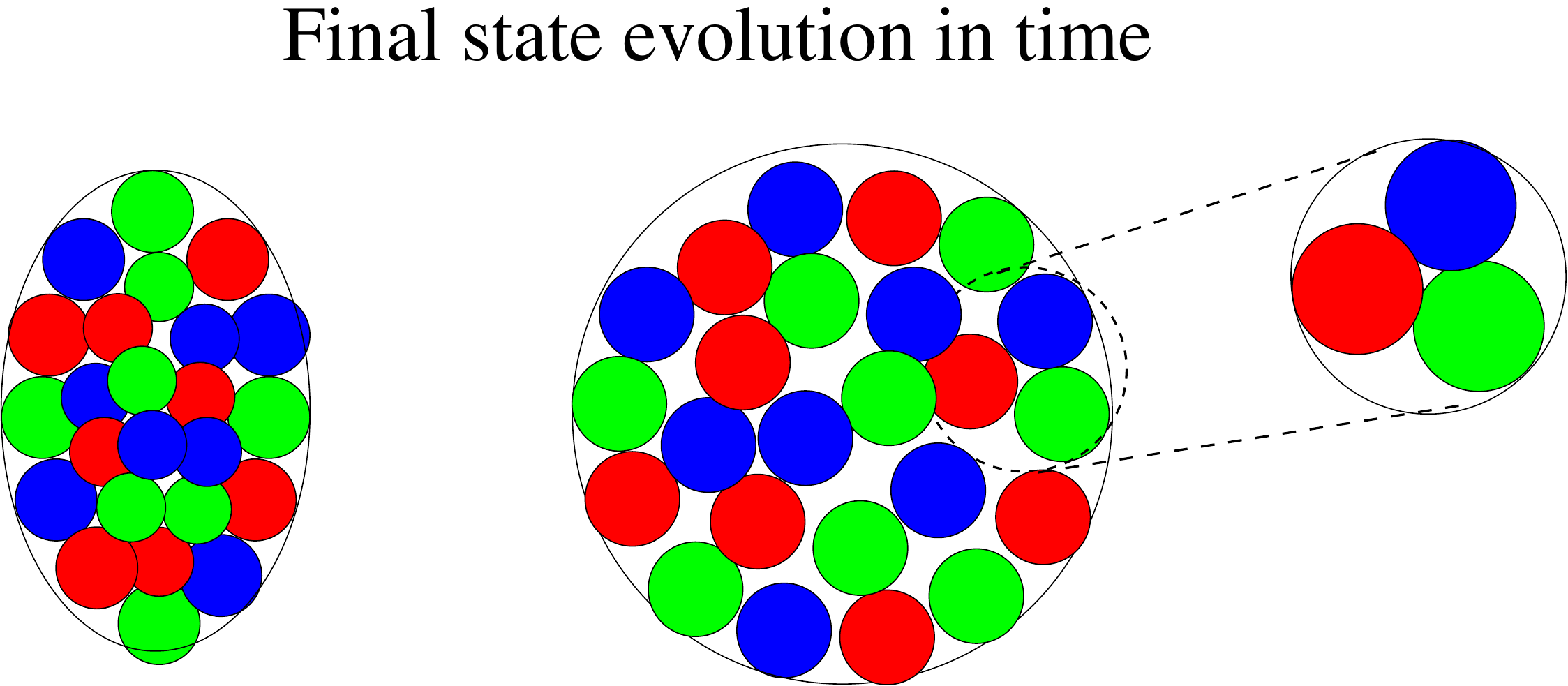}
    \caption{Final state evolution of the dense fields. Note that the number
      of dense fields is the same at all times. The figure is meant as a
      schematic illustration. Importantly, some of the fields will have
      anticolors (antiquarks) to form mesons and antibaryons.}
  \label{fig:final}
\end{center}
\end{figure}

Let us now look at the system in more details. The initial domain size are
fixed by the energy density, cf. Eq.~\ref{eq:edensity}; the domains cannot be
larger, even it would be energetically favorable, and conserve energy at the
same time. Now we assume that the number of domains is fixed during the
expansion to conserve entropy (reversibility). This assumption means that the
domains are valence-quark-like degrees of freedom, i.e., they grow 1-to-1 into
the valence quarks in the final state hadrons. As entropy is essentially
conserved, the expansion of the domain sizes must be
adiabatic. Figure~\ref{fig:final} illustrates the final state evolution of the
dense fields as a function of time. Let me be clear that the domain size at a
given rapidity in the initial and final state do not have to be related. In
both cases the important things is that domains are dense, which can have
different meaning in the initial and final state. In fact one would expect
that the domain sizes in the final state are larger because the smaller the
domains get in the initial state the less likely they will be to
interact. 

If $N_{\rm d}$ is constant then it means that the total final multiplicity,
$N$, is $N \propto \qd^3$. Furthermore, as a domain expands its energy (mass)
decreases. This difference in energy must therefore be converted to kinetic
energy so that for the final state total energy, $E$, one has $E \propto
\qd^4$. Let us now compare this to data from heavy-ion collisions. At the LHC
the \dndeta is approximately twice that at the maximum RHIC energy for the
most central collisions. To account for the difference in the longitudinal
direction it is estimated that $\dndeta \propto \qd^2$ (while $N \propto
\qd^3$). As the initial overlap region is supposedly the same at the LHC and
RHIC, this suggest that initial domains are ${\approx}40\,\%$ smaller at the
LHC than at RHIC. The transverse energy (2D) at midrapidity is for similar
reasons expected to scale with $\qd^3$. The dense field model presented here
suggests that transverse energy should therefore increase by ${\approx}40\,\%$
more than \dndeta going from RHIC to the LHC. The increase reported by CMS at
\snnt{2.76} was ${\approx}42\pm 15\,\%$~\cite{Chatrchyan:2012mb} while the
same increase measured by ALICE was found to be ${\approx}18 \pm
12\,\%$~\cite{Adam:2016thv}. Recent preliminary results from ALICE for
\snnt{5.02}, where the lever arm is longer, seem to suggest that the increase
is likely closer to the top end of the ALICE limit.

Studies of \dndeta and \et as a function of beam energy at
RHIC~\cite{Adler:2004zn} in general show a smaller rise than suggested by the
dense field picture. One could suspect that the reason for this is that
collisions at lower energy are only semi-transparent, meaning that significant
baryon number is transported to midrapidity, which biases the \et per
particle.\\

Finally, let me first give some thoughts on equilibration and thermalization
and then discuss why the dense fields would repel each other. For the final
state to behave as a medium, i.e., for a collective expansion to take place,
it must be required that a domain is formed so it can act as a coherent entity
and that the same is true for its neighbors. As the initial domains are
assumed to be coherent objects and as the final state formation merely
requires these domains to form a dense final state, this suggests that the
collective expansion can take place for early times/small collisional
systems. As previously noted, the scaling of energy density implies $\qd
\propto T$, but as the domain size does not require a full system in
equilibrium one could think of the proposed dense fields as useful structures
for describing the pre-equilibrium physics, e.g., for Debye screening where
one expects effects on scales of $1/T$.

Now let us go on to the bigger problem of why the domains would repel each
other. In the CGC, the gluons are treated as stochastically distributed
sources but realistic correlations between gluons have been studied in CGC
like models~\cite{Mueller:2002pi,Iancu:2002aq}. Importantly, in both works
they found that the state is color neutral for gluon wavelengths larger than
$1/\qs$ due to active shielding from other gluons (while color appears
randomly distributed for wavelengths less than $1/\qs$). If the same kind of
shielding takes place for the proposed color domains then this would explain
why the domains are stable and also suggest that they interact in a similar
way as magnetic field lines repelled by a superconducting magnet. This seem to
have some relation to the Debye screening length, not as a screening caused by
uncorrelated high color charge densities but rather as an active screening
process enforcing color neutrality on scales larger than $1/\qd$.

\subsection{Jet quenching in the QGP}

Let us try to do a qualitative estimate of how jet quenching depends on $\qd$
for a static source. If one considers a purely geometrical cross section then
one expects $\sigma \propto 1/\qd^2$ and this is also somewhat motivated from
calculations of valence quarks interacting with a
CGC~\cite{Dumitru:2002qt}. The energy loss in each interaction, $\delta E$, is
supposedly also of order $\qd$. For incoherent scatterings one therefore
obtains the total energy loss
\begin{eqnarray}
  \Delta E & \propto & \rho L \sigma \delta E \\
           & \propto & \qd^3 L \frac{1}{\qd^2} \qd \\
           & \propto & \qd^2 L \\ 
           & \propto & \sqrt{\varepsilon} L \\ 
           & \propto & T^2 L, 
\end{eqnarray}
where $L$ is the path length.

This rough estimate has several caveats as it does not take into account the
expansion of the medium as the probe traverses it or the effect of the initial
longitudinal momentum. It is still useful to highlight that, in the same way
as saturation limits the particle production in the initial state, the dense
fields in the final state will limit the quenching and make it rise less than
linear with the energy density. Furthermore, it is interesting that the scale
of the dense fields, \qd, naturally couples to the idea of the medium being
unable to resolve the jet structure so that the jet is quenched
coherently~\cite{CasalderreySolana:2012ef}.
 
\section{Discussion}
\label{sec:discussion}

In this section the two main ideas of the paper will be discussed in a broader
context. \\

The first idea is centered around the remarkable observation that many
phenomena we observe in the QGP are nearly reversible. This is by no mean a
trivial observation as the prediscovery expectations for the QGP was that it
would be weakly coupled. There were several reasons to expect this: asymptotic
freedom, lattice QCD energy densities in the QGP were consistent with
expectations for a relativistic quark-gluon gas, and the general result that
the free energy is eventually dominated by entropy at large temperatures.
Similarly, if Hadronization would proceed via very heavy hadronic states, as
is the underlying idea of the Hagedorn limiting temperature, then there would
be a clear separation between the QGP and hadronic phases.\\ This conflict
between expectations and observations points to a fundamental lack in our
understanding of QCD. The proposed solution in this paper is a duality between
QGP and hadrons that would give some insights into QGP properties: QGP and
hadrons are fundamentally strongly coupled, the expansion is driven by
asymptotic freedom, and the phase transition is driven by the energy density
cf.\ Eq.~\ref{eq:energy}.

In the second half of the paper, the idea of Universal color domain-like
degrees of freedom was presented. Here, it will be outlined how a full
generator building on these ideas could look. The initial state would involve
CGC-like calculations of domain scatterings followed by a non-CGC requirement
that the scattered domains are dense in the final state (similar to the idea
of the EKRT model~\cite{Eskola:1999fc}). Similarly to CGC longitudinal glasma
fields~\cite{Gelis:2010nm} and Lund string models as implemented in
PYTHIA~\cite{Sjostrand:2007gs}, there would have to be structures that are
long range in rapidity to get long range correlations. These intermediate long
range structures does not have to be dense fields as long as they break up
early to form the QGP (dense field phase). To get the proper strangeness
scaling with multiplicity~\cite{ALICE:2017jyt} one could consider to use color
ropes for the initial ``strings''~\cite{Bierlich:2014xba}. One would need to
implement something like the \pp core-corona model in
EPOS~\cite{Pierog:2013ria} to separate the hard (\ee like) and soft (QGP)
components. The microscopic implementation of the hydrodynamics is very
similar to the ``shoving'' model~\cite{Bierlich:2016vgw}. Note that the basic
picture of the expansion in the domain model is similar to the experiments
done with strongly interacting lithium atoms~\cite{Cao:2010wa} because the order of the domains must to first order be
preserved for the liquid to be perfect. Hadronization will proceed via
recombination of the domains as they achieve the size \qh.

\section{Summary}

In this paper we have tried to speculate on QCD implications of mini-QGP
formation in proton-proton collisions. It has been proposed that, due to the
reversible nature of the QGP, time reversal could be a powerful tool to
understand the relationship between the QGP and hadrons. We have tried to
extend the CGC ideas to the full initial and final state as a dense field
picture. This QGP-hadron duality would suggest that there should be relations
between initial and final state \pp physics. It is the hope of the author,
that better qualitative and quantitative tests can be devised to validate or
falsify the proposed picture.

Finally, a goal for me in pursuing such a picture has been that it should, if
correct, help explain to non-physicists why we collide heavy-ions: \\ Each
nucleon contains a snapshot of how the Universe looked a few microseconds
after the Big Bang. In this way we are not only children of the stars but
carries within ourselves also an imprint of the Big Bang itself. When we
collide nuclei, they replay for us the movie of creation.

\end{document}